# SUCROSE SOLUTIONS AS PROSPECTIVE MEDIUM TO STUDY THE VESICLE STRUCTURE: SAXS AND SANS STUDY

M.A. Kiselev*, P. Lesieur#, A.M. Kisselev*, D. Lombardo#, M. Killany*, S. Lesieur%

\* - Frank Laboratory of Neutron Physics, JINR, 141980 Dubna, Moscow reg., Russia
\# - LURE, Bat. 209-D, B.P. 34, F-91898 Orsay, France
% - Pharmaceutical Faculty, University Paris-Sud, Chatenay Malabry F – 92296, France

**ABSTRACT**

The possibility to use sucrose solutions as medium for X-ray and neutron small-angle scattering experiments has been explored for dimyristoylphosphatidylcholine (DMPC) vesicles and mixed DMPC/$C_{12}E_8$ aggregates. The influence of sucrose concentration on phospholipid vesicles size and polydispersity has been investigated by complimentary X-ray and neutron scattering. Sucrose solutions decreased vesicle size and polydispersity and increased a contrast between phospholipid membrane and bulk solvent sufficiently for X-rays. 40% sucrose in $H_2O$ increased X-ray contrast by up to 10 times compared to pure $H_2O$. The range of sucrose concentration 30%-40% created the best experimental conditions for the X-ray small-angle experiment with phospholipid vesicles.

*Correspondence to* :
M. A. Kiselev - Frank Laboratory of Neutron Physics, Joint Institute for Nuclear Research, 141980 Dubna, Moscow region, Russia.
E-mail: kiselev@nf.jinr.ru, fax, 7-096-21-65882; telephone: 7-096-21-66977.

# 1. INTRODUCTION

Vesicles are very important in many different areas of science and technology. They serve as delivery agents for drugs, genetic materials and enzymes through living cell membrane and other hydrophobic barriers in pharmacology, medicine, genetic engineering, cosmetic and food industry [1,2]. Vesicle size appears to be a major factor in its permeation through tumor microvessels and its local residence in tumor tissue. A vesicle size of 1000Å may be a pivotal size of vesicles for tumor targeting and long vesicle circulation in blood [3].

The small-angle X-ray (SAXS) and neutron (SANS) scattering experiment on vesicles from phospholipids serving as a model biological membranes is a useful method for determination of membrane and vesicle structure [4-9]. The Guinier approximation of SANS spectra was used for the calculation of membrane thickness of egg yolk phosphatidylcholine vesicles ($d_l$=41.7Å at 20°C), dipalmitoylphosphatidylcholine (DPPC) vesicles ($d_l$=49.6Å at 20°C), and mixed lipid/surfactant vesicles [6-8]. A strip-function model of evaluating SANS curves was developed recently which allows to determine important membrane parameters describing the geometry and hydration of large oligolamellar vesicles [9]. The radius of large unilamellar vesicles (R≥1000 Å) at itself cannot be determined from SANS or SAXS scattering experiment due to the technical limitation of modern spectrometers and large polydispersity of vesicles in size and morphology [9-11]. The methods to create small unilamellar vesicles (SUVs) with low polydispersity such as fast dilution or temperature jump protocol were developed for mixed lipid/surfactant systems [5,12]. The problem of creation of the SUVs with low polydispersity from a pure lipid system without surfactant are not solved till now.

The application of X-ray small-angle scattering to the study of vesicular systems has a limitation due to a weak contrast between phospholipid bilayer and water. Sucrose solutions were used for the contrast variation make X-ray scattering of myoglobin comparable to the neutron one [13]. The interaction of sugars with phospholipid molecules is not clear. The experimental results obtained with multilamellar vesicles (MLVs) and unilamelar vesicles (ULVs) are in contradiction. Trehalose does not influence the main transition temperature $T_m$ of DMPC ULVs, but with MLVs the $T_m$ is increased by the presence of trehalose [14]. For DMPC MLVs, sucrose causes a slight increase of $T_m$ by the value less than 2°C for sucrose concentration from 0 up to 69% [15].

In our previous study, the 40% sucrose solution was used for the investigation of micelle to vesicle transition in the mixed DMPC/ionic surfactant system via SAXS [5]. In the present study, the possibility to use sucrose solutions as medium for X-ray and neutron small-angle scattering experiment was explored for DMPC vesicles and mixed DMPC/nonionic surfactant aggregates. The influence of sucrose concentration on phospholipid vesicles size and polydispersity has been investigated by complimentary X-ray and neutron scattering.

## 2. MATERIALS AND SAMPLE PREPARATION

**D**imyristoylphosphatidylcholine (DMPC) - $C_{36}H_{72}NO_8P$, 8 Lauryl Ether ($C_{12}E_8$) - $C_{25}H_{58}O_9$, sucrose - $C_{12}H_{22}O_{11}$ were purchased from Sigma (France), $D_2O$ (99.9% deuteration) from Isotop (Russia). $H_2O$ was of Millipore standard (18MΩ·cm). The 1% (w/w) of DMPC was dissolved in sucrose/water solution. The samples were heated to a temperature above the main phase transition ($T_m$=23°C) and then cooled down to 10°C. The cooling-heating cycle was repeated 5 times with sample shaking. MLVs were created under this process. For the formation of unilamellar vesicles the obtained suspension of MLVs was passed 25 times through polycarbonate filter with 500Å pore diameter of LiposoFast Basic extruder (Avestin, Canada) [16].

## 3. METHODS

The SAXS measurements were carried out at D22 spectrometer of DCI synchrotron ring at LURE, France, and SANS measurements - at YuMO spectrometer of pulse neutron reactor IBR-2 at Dubna, Russia [17,18].

The model of infinitely thin sphere was applied to interpret the SAXS curves in the region of scattering vector q from 0.005Å$^{-1}$ to 0.04Å$^{-1}$ [5]. The macroscopic cross section of vesicles in this model is given by

$$\frac{d\Sigma}{d\Omega}(q) = n \cdot (4\pi \cdot R^2 \cdot d_l \cdot \Delta\rho)^2 \cdot \left(\frac{Sin(qR)}{qR}\right)^2, \qquad (1)$$

where n is the number of vesicles per unit volume, R-vesicle radius, $d_l$- membrane thickness, $\Delta\rho$ - X-ray contrast between membrane and aqueous sucrose solution. Due to the vesicles polydispersity, the macroscopic cross section was convoluted with introduced Gaussian distribution of vesicles radii, which gave the final expression for the vesicles macroscopic cross section $I_c(q)$

$$I_c(q) = \frac{\int_{R_o-3\sigma}^{R_o+3\sigma} \frac{d\Sigma}{d\Omega}(q,R) \cdot \exp\left(-\frac{(R-\bar{R})^2}{2\cdot\sigma^2}\right) \cdot dR}{\int_{R_o-3\sigma}^{R_o+3\sigma} \exp\left(-\frac{(R-\bar{R})^2}{2\cdot\sigma^2}\right) \cdot dR}, \qquad (2)$$

where $\sigma$- standard deviation of R and $\bar{R}$ - average radius of vesicles [18].

The model of hollow sphere was applied for the interpretation of the SANS curves in the region of scattering vector q from 0.007Å$^{-1}$ to 0.14Å$^{-1}$ [6, 18]. In this approach the macroscopic cross section of vesicles can be written as

$$\frac{d\Sigma}{d\Omega}(q) = n \cdot (\Delta\rho)^2 (4\pi/q^3)^2 (A_2 - A_1)^2, \text{ with } A_i = Sin(qR_i) - (qR_i)Cos(qR_i) \qquad (3)$$

where, n is the number of vesicles per unit volume, $\Delta\rho$- contrast for neutrons, $R_1$ and $R_2$ correspond to the inner and outer radii of the vesicle, respectively. The bilayer thickness is then $d_l = R_2 - R_1$. Two types of vesicle distribution were used to describe the SANS curves. The symmetrical Gaussian distribution as in Eq. (2) and nonsymmetrical Schulz distribution

$$G(R) = \frac{R^m}{m!} \cdot \left(\frac{m+1}{\bar{R}}\right)^{m+1} \cdot \exp\left[-\frac{(m+1)\cdot R}{\bar{R}}\right] \qquad (4)$$

where, m≥1 is integer [9]. The macroscopic cross section $I_c(q)$ was calculated in this case via convolution of Eqs. (3) and (4) from $R_{min}$=70Å to $R_{max}$=2000Å. The $I_c(q)$ value was corrected by the resolution function of YuMO spectrometer as described in [17].

In order to calculate the model parameters (R, $d_l$, $\Delta\rho$, m, $\sigma$), the experimentally measured X-ray and neutron macroscopic cross section were fitted by least-square minimization with calculated macroscopic cross section $I_c(q)$. Polydispersity P of vesicles were characterized by the half width at the half height (HWHH) of distribution function (4). For the Gaussian distribution $P=1.118\cdot\sigma/\bar{R}$.

## 4. RESULTS AND DISCUSSION
### 4.1. SANS study

The scattering cross sections of DMPC vesicles were measured in pure $D_2O$ and in $D_2O$ with 5, 10, and 20% sucrose (w/w) at T=30°C (liquid $L_\alpha$ phase of DMPC) [15]. The experimental spectra were fitted with Schulz distribution function according to Eqs. (3) and (4). Results of calculations are presented at Table 1. The increase of sucrose concentration in the solution decreases the average vesicles radius from the value of 260Å at 0% sucrose to the value of 200Å at 20% sucrose and decreases the vesicles polydispersity from the value of 37% to the value of 27%, respectively. The membrane thickness $d_l$ is not influenced by sucrose with the accuracy of applied method. Additionally, the spectrum from vesicles in the 20% sucrose solution was fitted with Gaussian distribution function. The experimental spectra and fitted curves for the DMPC vesicles in $D_2O$ with 20% sucrose are presented in Fig. 1. The fitted curves with Schulz and Gaussian distribution functions practically coincide. The results of application of the hollow sphere model with the Gaussian distribution give the following values: $\bar{R}$=190±5Å, $d_l$=34.8±0.5Å, $\sigma$=50±5Å, P=31%, which are in agreement with results obtained with Schulz distribution function, except the value of polydispersity. The application of Gaussian distribution gives polydispersity of 31% that is larger than the value of 27% obtained with the Schulz distribution function.

The measurements of the membrane thickness of LUVs at T=10°C (gel $L_{\beta'}$-phase of DMPC) were carried out to check conclusion about permanent DMPC membrane thickness in the liquid $L_\alpha$ phase. The LUVs were prepared by extrusion through 1000Å pores [16]. The SANS measurements were carried out in the sucrose/$D_2O$ solutions with 0%, 20%, 30%, 40% sucrose concentrations. The membrane thickness $d_l$ was calculated from the slope of Kratky-Porod plot according to the Guinier approximation as described in [7-8, 18, 22]. The results of calculations are presented in Fig. 2. The decrease of DMPC membrane thickness $d_l$ at T=10°C by the value of 1.4±1.8Å upon the increase of sucrose concentration by 40% is within the interval of experimental errors and can be explained as a result of the contrast decrease. The decrease of the contrast leads to the underestimation of the membrane thickness in the Guinier approximation [19, 20]. The same Guinier approximation was applied to calculate the DMPC membrane thickness in the liquid $L_\alpha$ phase at T=30°C. $d_l$=38.8±0.8Å was evaluated from the Kratky-Porod plot for DMPC vesicles in $D_2O$. The $d_l$ calculated from the Kratky-Porod plot is by 3.6±1.0Å larger than the $d_l$ calculated from the hollow sphere model, $d_l$=35.2±0.2Å. The choice between two different values of the $d_l$ will be made in the SAXS study section.

DMPC membrane thickness is not influenced by sucrose in the range of sucrose concentration from 0% to 40%. It is an important fact which gives us opportunity to consider sucrose as prospective medium for SAXS and SANS experiments with phospholipid vesicles.

### 4.2. SAXS study

Phospholipid membranes are objects with a very poor contrast for X-ray. Under the assumption that the structure of lipid membrane is not perturbed by the addition of sucrose and that no specific adsorption of sucrose or water occurs at the membrane surface, the X-ray scattering length density of DMPC molecule in the $L_\alpha$-phase $\rho_{DMPC}= 0.957 \cdot 10^{11}$ cm$^{-2}$. The X-ray scattering length density of sucrose solution is calculated by the expression

$$\rho_{solvent} = 16.979 \cdot \left(\frac{1-\chi}{\chi} + 0.0532 \cdot \chi\right) \cdot D(\chi) \cdot 10^{11}, cm^{-2} \qquad (5)$$

where, $\chi$- sucrose concentration (w/w), $D(\chi)$- density of sucrose/water solution, g/cm$^3$ [21]. X-ray contrast between membrane and solvent is determined as $\Delta\rho = \rho_{DMPC} - \rho_{solvent}$.

Fig. 3 presents the calculated value of $\rho_{solvent}$ at T=30°C in the range of sucrose concentrations from 0 to 60% and the values of $\rho$ for two types of phospholipids in liquid $L_\alpha$-phase: DMPC and DPPC. The contrast for DMPC molecule in pure water is positive and for DPPC molecule is negative, but in any case the absolute value of contrast is about $0.14 \cdot 10^{10}$ cm$^{-2}$. The absolute value of contrast $|\Delta\rho|$ for sucrose solution increases sufficiently with the increase of sucrose concentration. 40% sucrose in H$_2$O increase X-ray contrast by up to 10 times compared to pure H$_2$O. The macroscopic cross section of DMPC vesicles at T=30°C increases with the increase of sucrose concentration due to the increase in the contrast as shown in Fig. 4. At sucrose concentration 15% the statistical errors decreased down to the values that give opportunity for the model application. The results of the application of the infinitely thin sphere model (Eqs. (1) and (2)) to the experimental spectra and the calculated by Eq. (5) values of contrast are presented in Table 2. These theoretical values of contrast were compared with the experimentally determined values of contrast, that is calculated from the measured values of macroscopic cross section, $\bar{R}$, $\sigma$, and $d_l$, according to the Eqs. (1) and (2). The $d_l$ was determined from the SANS experiment. SANS results at T=30°C gave two different values for the membrane thickness:. The application of $d_l$=38.8Å for the calculation of the experimental value of the contrast gave good coincidence with the theoretical value. The application of $d_l$=35.2Å leads to the disagreement between theory and experiment. Consequently, the contrast analysis gave argument to the rights of $d_l$=38.8Å±0.8Å for DMPC membrane thickness at T=30°C.

As it is seen from Table 2, the discrepancy between calculated and measured contrast is no greater than 20%, except for big difference at 50% sucrose. The big error for 50% sucrose arises from the uncertainty in the value of vesicle radius which is 130±105Å. The 20% difference between calculated and measured values of the contrast for all sucrose concentrations in the range of 15%-45% are in agreement with the minimum value of polydispersity, 33%. The average radius of vesicles and polydispersity have the minimum values in the range of sucrose concentrations 30-40%. This interval exhibits the best experimental conditions for SAXS. Figs. 5 and 6 demonstrate this conclusion for the case of micellar to lamellar transitions in the DMPC/C$_{12}$E$_8$ system [8]. This system possesses a property of supramolecular aggregates transformation from the micelles phase to lamellar phase upon the temperature alteration [22, 23]. The DMPC/C$_{12}$E$_8$ system was studied by SANS in D$_2$O (Fig. 5) and by SAXS in 40% sucrose solution (Fig. 6). The macroscopic cross section for neutrons and X-rays can be compared at T=10°C. The lines extrapolating to the zero value of the scattering vector give $d\Sigma/d\Omega(0)$=0.6cm$^{-1}$ for X-ray scattering and $d\Sigma/d\Omega(0)$=9.9cm$^{-1}$ for neutron scattering. The DMPC concentration in neutron experiment is two times smaller of the DMPC concentration in X-ray experiment. Consequently, the normalized to the DMPC quantity macroscopic cross sections of neutrons and X-rays differ by a factor 33. Taking into account the high flux of X-rays at the synchrotron source, we can conclude that the introduction of sucrose solutions for X-ray scattering gives comparable experimental conditions for the use of D$_2$O for neutron scattering.

5. CONCLUSIONS

Neutron and X-ray small-angle scattering are powerful complementary tools for the investigation of physicochemical systems of biological interest like model membranes and mixed lipid/surfactant systems. The neutron scattering is advantageous when partly or fully deuterated molecules can be synthesized and used as deuterated labels in the sample. For non-deuterated materials, resolution and flux of the X-ray scattering at a synchrotron source are of extreme interest.

Sucrose solutions have three advantages. First is the possibility to create a monodispersed population of vesicles, which sufficiently improves the experimental conditions to study vesicles structure. Second is the increase in intensity, which is large enough to allow for the study of the structure of diluted aggregates by SAXS. Third is the constant membrane thickness at sucrose concentration below 40%. Sucrose/water solutions are prospective media for the SAXS and SANS application to the investigation of vesicles structure and structure of mixed lipid/surfactant systems. The range of sucrose concentration 30%-40% create the best experimental conditions for the X-ray small-angle experiment with phospholipid vesicles.

ACKNOWLEDGEMENTS
This study was supported by the TMR program for Great Instruments. The authors are grateful to Dr. P.Balgavy (Faculty of Pharmacy, J.A. Comenious University, Bratislava) and to Dr. M. Ollivon (URA 1218 of CNRS, France) for the fruitful discussions.

REFERENCES
[1] D.D. Lasic, in Ed. R. Lipowsky, E. Sackmann (Ed), Hadbook of Biological Physics, vol.1 , Elsevier Science B.V, 1995, pp. 491-519.
[2] M. Rosoff, Vesicles, Marcel Dekker, Inc. 1996.
[3] A. Nagayasu, K. Uchiyama, H. Kiwada, Adv. Drug Delivery Rev. 40 (1999) 75-87.
[4] P. Lagner, A.M. Gotto, J.D. Morrisett, Biochemistry 18 (1979) 164-171.
[5] P. Lesieur, M.A. Kiselev, L.I. Barsukov, D. Lombardo, J. Appl. Cryst. 33 (2000) 623-627.
[6] P. Balgavy, M. Dubnichkova, D. Uhrikova, S. Yaradaikin, M. Kiselev, V.Gordeliy, Acta Phys. Slovaca 48 (1998) 509-533.
[7] M.A. Kiselev, P. Lesieur, A.M. Kisselev, C. Grabiel-Madelmond, M. Ollivon, J. Alloys and Compounds 286 (1999) 195-202.
[8] T. Gutberlet, M. Kiselev, H. Heerklotz, G. Klose, Physica B 381-383 (2000) 276-278.
[9] H. Schmiedel, P. Joerchel, M. Kiselev, G. Klose, Accepted in J. Phys. Chem. 2000.
[10] B.L. Mui, P.R. Cullis, E.A. Evans, T.D. Madden, Biophys. J. 64 (1993) 443-453.
[11] D.G. Hunter, B.J. Frisken, Biophys. J. 74 (1998) 2996-3002.
[12] P. Schurtenberger, N. Mazer, W. Kanzig, J. Phys. Chem. 89 (1985) 1042-1049.
[13] K. Ibel, H.B. Stuhrmann, J. Mol. Biol. 93 (1975) 255-265.
[14] J.H. Crowwe, L.M. Crowe, J.F. Carpenter, A.S. Rudolph, A.A. Wistrom, B.J. Spargo, T.J. Anchordogy, Biochim. Biophys. Acta 947 (1988) 367-384.
[15] C.H. J.P. Fabrie, B. de Kruijff, J. de Gier, Biochim. Biophys. Acta 1024 (1990) 380-384.
[16] R. C. MacDonald, R. I. MacDonald, B. P. Menco, K. Takeshita, N. K. Subbarao, L. R. Hu, Biochim. Biophys. Acta 1061 (1991) 297-303.
[17] Y.M. Ostanevich, Macromol. Chem., Macromol. Symp. 15 (1988) 91-103.
[18] L. A. Feigin, D. I. Svergun, Structure Analysis by Small-Angle X-Ray and Neutron Scattering, Plenum Publishing Corporation, New York, 1987.
[19] M.A. Kiselev, A.M. Kisselev, S. Borbely, P. Lesieur, JINR publication E17, 19-98-305 (1998) 81-86.
[20] N. Gorski, Y. Ostanevich, Ber. Bunsenges, Phys. Chem. 94 (1990) 734-741.
[21] E.J. Barber, Nat. Cancer Inst. Monogr. 21 (1966) 219-239.

[22] M.A. Kiselev, P. Lesieur, A.M. Kisselev, S.A. Kutuzov, L.I. Barsukov, T.N. Simonova, T. Gutberlet, G. Klose, JINR publication E3,14-98-168 (1998) 52-57.
[23] M.A. Kiselev, P. Lesieur, D. Lombardo, A.M. Kisselev, T. Gutberlet, Chem.&Phys. of Lipids 107 (2000) 72.

Table 1.

The sucrose influence on the DMPC vesicle structure at T=30°C. The model of hollow sphere with Schulz distribution. m - integer in functions (4), $\bar{R}$ - average value of vesicles radii, $d_l = R_2-R_1$ - membrane thickness, P - polydispersity (half width at the half height of the distribution function)

| Sucrose concentration (w/w) | m | $\bar{R}$, Å | $d_l$, Å | P,% |
|---|---|---|---|---|
| 0 | 8 | 260±5 | 35.2±0.2 | 37 |
| 5 | 11 | 235±5 | 34.8±0.5 | 32 |
| 10 | 11 | 220±5 | 34.8±0.5 | 32 |
| 20 | 18 | 200±5 | 34.8±0.2 | 27 |

Table 2.

The sucrose influence on the DMPC vesicle structure at T=30°C. The parameters of spherical model with infinitely thin surface. $\bar{R}$ - average value of vesicles radii, σ-standard deviation of R. The theoretical value of $(\Delta\rho)^2$ is calculated from the electron density difference according to the Eq. (5). The experimentally measured value $(\Delta\rho)^2$ is calculated according to the Eqs. (1) and (2). P=1.18·σ/R - is characterized the value of polydispersity.

| Sucrose concentration, % | $\bar{R}$, Å | σ, Å | P, % | $(\Delta\rho)^2$, $10^{20}$cm$^{-4}$ theory | $(\Delta\rho)^2$, $10^{20}$cm$^{-4}$ experiment |
|---|---|---|---|---|---|
| 15 | 228±2 | 83±2 | 42 | 0.16 | 0.13 |
| 20 | 237±1 | 74±1 | 37 | 0.28 | 0.33 |
| 25 | 213±1 | 78±1 | 39 | 0.62 | 0.57 |
| 30 | 218±1 | 61± | 33 | 0.91 | 1.14 |
| 35 | 216±1 | 60±1 | 33 | 1.44 | 1.49 |
| 40 | 218±1 | 61±1 | 33 | 1.88 | 2.34 |
| 45 | 216±1 | 77±1 | 43 | 2.55 | 2.62 |
| 50 | 130±10 | 105±2 | 96 | 3.19 | 5.54 |

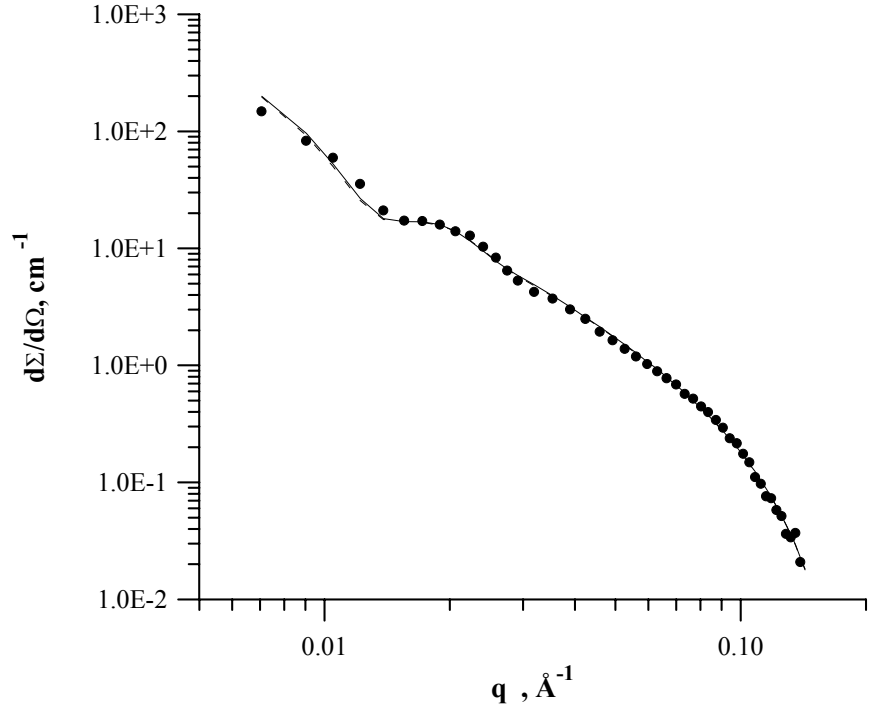

Figure 1. SANS curve from DMPC vesicles in D$_2$O with 20% sucrose, T=30°C (points). Fitted curves: solid line - Gaussian distribution ($\bar{R}$=190±5Å, d$_l$=34.8±0.5Å, σ=50±5Å, P=31%), dashed line - Schultz distribution ($\bar{R}$=200±5Å, d$_l$=34.8±0.2Å, m=18, P=27%).

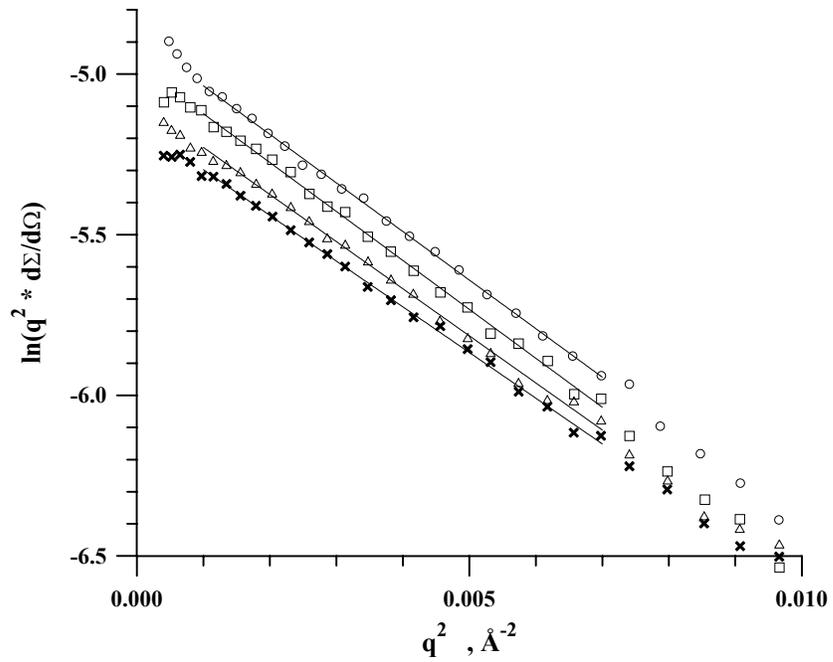

Figure 2. Kratky-Porod plot for DMPC vesicles at T=10°C. $D_2O$, $d_l$=44.5Å±1.0Å (circles); 20% sucrose solution, $d_l$=44.7±0.8Å (squares); 30% sucrose solution, $d_l$=43.7±0.8Å (triangles); and 40% sucrose solution, $d_l$=43.1±0.8Å (crosses).

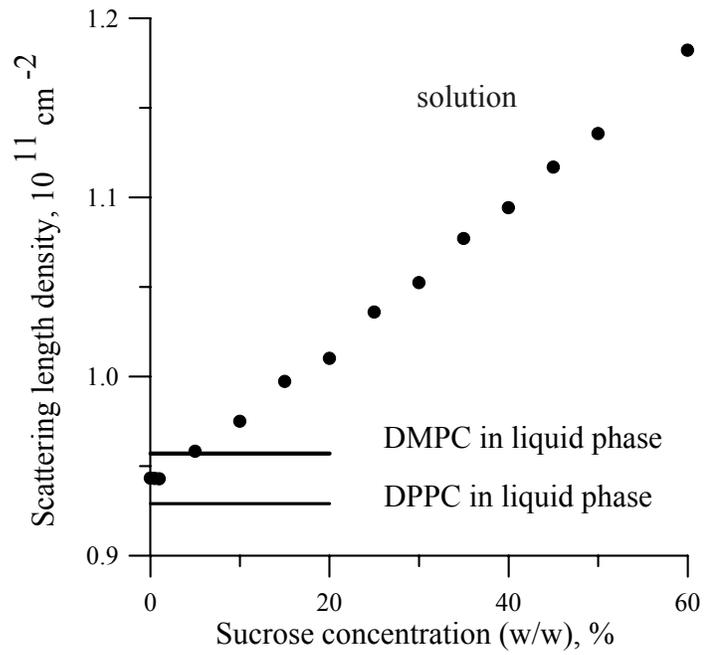

Figure 3 . X-ray scattering length density $\rho$ of aqueous sucrose solution at T=30°C as function of sucrose concentration (points). The two lines indicates the value of $\rho$ for DMPC and DPPC membrane in the liquid $L_\alpha$-phase.

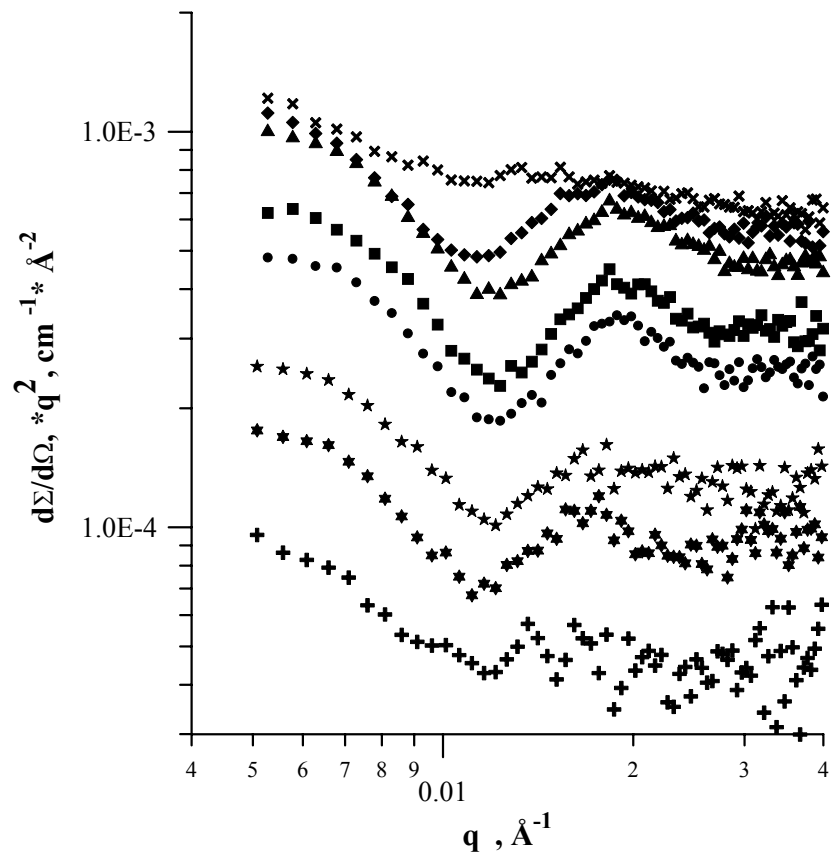

Figure 4. SAXS curves for DMPC vesicles in the aqueous sucrose solutions at T=30°C. The macroscopic cross section increase with increase of sucrose concentration, 15% (crosses), 20% (6 angles stars), 25% (5 angles stars), 30% (circles), 35% (squares) 40% (triangles), 45% (rhombuses), 50% (crosses).

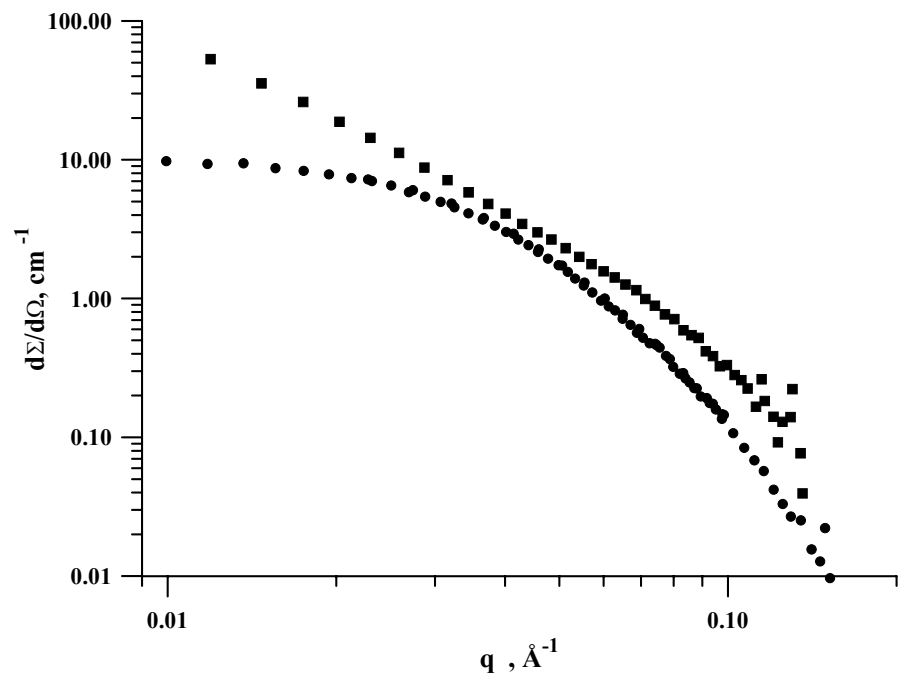

Fig. 5. SANS spectra from 7.5mM DMPC/4mM $C_{12}E_8$ systems at T=10°C (points) and T=38°C (squares). Mixed lipid/surfactant aggregates transform from micelles with $R_g$=33Å at T=10°C to the unilamellar bilayers and the vesicles with a membrane thickness of 37.7Å at T=38°C.

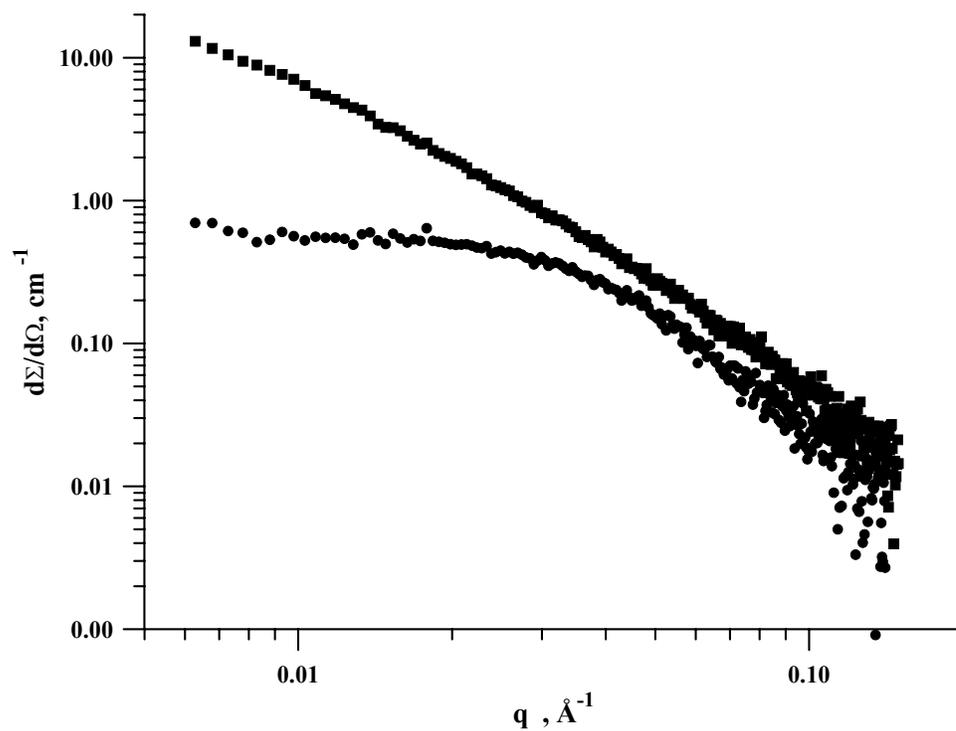

Fig. 6. SAXS spectra from 15mM DMPC/11.25mM $C_{12}E_8$ system in 40% sucrose solution at T=10°C (points) and T=50°C (squares). Mixed lipid/ surfactant aggregates transform from micelles with $R_g$=38.5Å at T=10°C to the unilamellar bilayers and the vesicles at T=50°C.